\documentclass[runningheads,citeauthoryear]{apinv38}
\usepackage{epsfig,cite,graphics}
\usepackage[bulgarian,english]{babel}

\begin{document}
\title{Macro-flickering of AQ~Mensae \\ on the daily time-scales: \\ Parameters and quasi-period modes}
\author{Ts.~B.~Georgiev$^1$~, R.~K.~ Zamanov $^1$~, S.~Y.~Stefanov$^{1,2}$~}

\titlerunning{Macro-flickering of AQ~Men}
\authorrunning{Georgiev, Zamanov, Stefanov }
\tocauthor{Ts.~B.~Georgiev, R.~K.~ Zamanov,   S.~Y.~Stefanov}
\institute {$^1$~Institute of Astronomy and National Astronomical Observatory, \\ Bulgarian Academy of Sciences, 72 Tsarigradsko Chaussee Blvd., 1784 Sofia, Bulgaria\\
$^2$~Department of Astronomy, Sofia University "St. Kliment Ohridski", \\James Boucher 5, BG 1164 Sofia, Bulgaria 
\newline \email{tsgeorg@astro.bas.bg } }
\papertype{Submitted on 27 March 2022; Accepted on 17 July 2022}	
\maketitle

\setcounter{page}{68}


\begin{abstract}
We analyzed $TESS$ photometric data of the flickering-active cataclysmic star AQ~Men in 2018--2019. We processed 7 sectors with 14 light curves (LCs) inside them, with a time resolution of 2 min. Aiming to study the "macro-flickering", with quasi periods (QPs) between 10 and 100 hours, we processed  LCs after 55 time reduced, with a time resolution of 1.83 hours. The method, developed earlier by us, includes comparing the LCs by their statistical and fractal parameters, as well as revealing QPs by minima of structure functions and relevant maxima of autocorrelation functions.

We distinguish the known high state of AQ~Men, in the sectors \#\# 01, 05, 08, 12, 13, as well as the low state, in sectors \#19, \#20. In the low state the LCs show noticeable eclipses with period 3.4 h, lower average fluxes, higher scatters, and additional QPs. By its statistical and fractal parameters, the macro-flickering of AQ~Men in the high state is similar to the ordinary flickering of 3 symbiotic binaries, studied by us earlier (see below). 

We found 92 QPs in the range of 20--70 h. We reveal 3 QP modes, at 20.9 h, 32.5 h and 54.1 h (1.149, 0.738 and 0.434 c/d; Fig.~7) within a standard error about 10\%. The last mode is the most populated one and seems to be a manifestation of the superorbital period. Other 4 QP modes of AQ~Men are added from the literature. The regularity of these 7 QP modes follows a power function with a base 1.57 and standard deviation 6.4\% (Fig.~8). This power model prognosticates 5 other QP modes: 3 internal and 2 external (Table 2). The bases of the power regularity models for the flickering of the symbiotic binaries RS~Oph, T~CrB, and MWC~560 (however in the time scale of minutes) are 1.55, 2.0, and 1.34, respectively (Table 1). For unknown reasons in these 4 cases we find (i) regularities with (ii) different bases. 

	\end{abstract}
.

\section*{Introduction}
AQ~Men (EC 05114-7955) is a nova-like cataclysmic variable star of 14.3-15.3 mag. Nova-like variables are systems in which no dwarf nova or classical nova outbursts have been recorded. They are binary systems in which a white dwarf primary is accreting material via Roche lobe overflow from a late-type main- sequence star. These systems have a high mass-transfer rate ($10^{-8} -10^{-9}$ M$_\odot$ per year) and the mass stream through the inner Lagrangian point forms an accretion disc. The properties of the different types of cataclysmic variables are described in the reviews by Warner (2003) and Knigge et al. (2011). 

AQ~Men was identified as a blue object in the Edinburgh-Cape Blue Object Survey (Stobie et al. 1995). Chen et al. (2001) suggested that this object could be a dwarf nova or a nova-like variable. Bruch (2020) noted that the absolute brightness of AQ~Men is too high for a dwarf nova, but is consistent with a faint nova-like variable.

\begin{figure}[!htb]  \begin{center} 
\centering{\epsfig{file=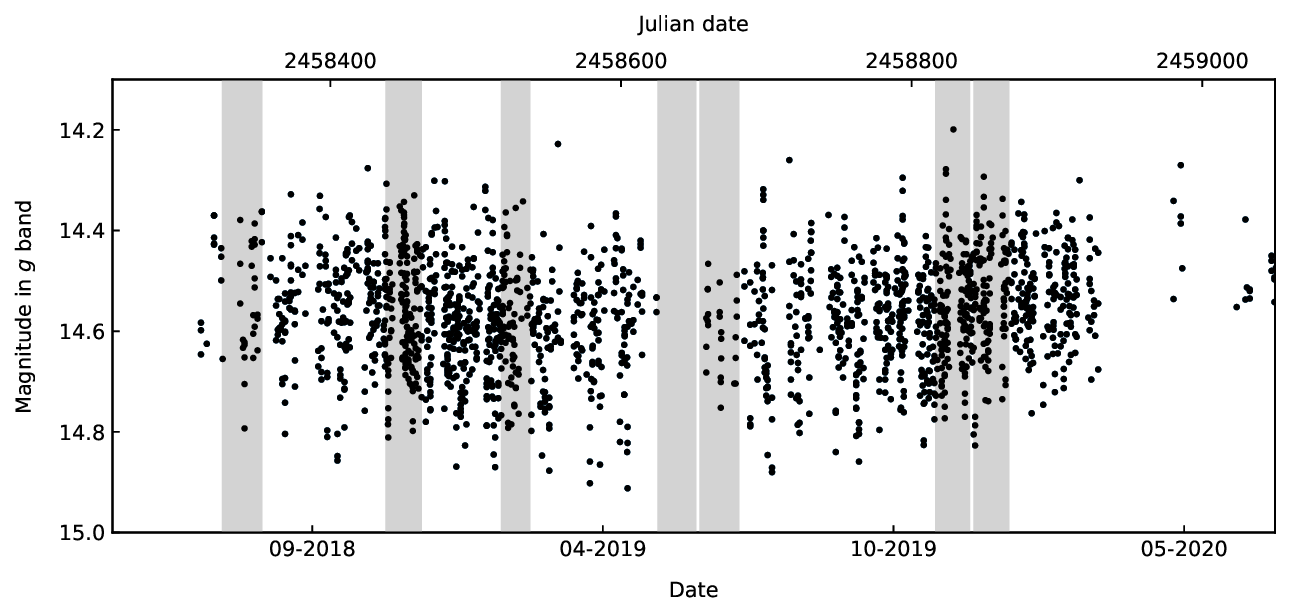, width=.9\textwidth}} 
\caption[] {ASAS-SN light curves of AQ~Men in the Sloan $g$ band. The photometry errors are smaller than the symbol size. The shaded regions represent the tines of $TESS$ sectors  \#\# 1,~5,~8,~12,`13,~19, and 20 in which AQ~Men was observed. The light curve in the present paper are fluxes in $(R+I)$ band approximately.}
\label{fig.1}   \end{center}  \end{figure}

I\l{}kiewicz et al. (2021) find in the light-curve of AQ~Men a low state which is visible at MJD about 56970 that lasted $\approx$100 days. The presence of a low state unambiguously classifies AQ~Men as a nova-like variable.

By analyzing radial velocities Chen et al.  (2001) derived an orbital period of 0.13 d (3.12 h). Later Armstrong et al. (2013) show that AQ~Men is an eclipsing system with an orbital period of 0.141 d (3.386 h). The eclipse depths are highly variable, which suggests that the eclipses are grazing.

I\l{}kiewicz et al. (2021) detected a tilted, precessing accretion disc in AQ Men. They show that the depths of the eclipses are changing with the orientation of the accretion disc and test some of the tilted accretion disc models. The precession period of the accretion disc is increasing during the TESS observations. However, it is still shorter than the period determined in the previous studies. They also detected a positive superhump that was unseen in the previous observations. This positive superhump has a strongly non-sinusoidal shape, which is not expected for a nova-like variable.

I\l{}kiewicz et al. (2021, Table 1) give the  frequencies of 24signals, detected in sector 13. They estimate a superorbital period of AQ~Men to be 57.02 h ($N=0.42093$~c/d).
 
In addition to the orbital period, Armstrong et al. (2013) detected a signal at 0.263 c/d (91.125 h) and a weaker signal at 7.332 c/d (3.273 h). These signals were interpreted as superorbital and negative superhump periods, respectively. In 2018 Bruch (2020, Figs.~11, 13) detected a strong signal at 8.25 d (198 h), as well as a weak signal at the orbital period of 3.901 h, and its half 1.955 h. 

The aims of the present paper are to reveal statistical and fractal parameters of the flux changes, as well as, to unveil quasi-periods (QPs) and their modes. This study is concentrated on daily time scales of 20--70 hours. While the "flickering" concerns the flux changes on the time scales tens of minutes, here we deal with macro-flickering on the scale of tens of hours. The flickering with period of  54--55 h (Fig.~7) seems to correspond to the superorbital period.  

\begin{figure*}
\begin{center}  
\centering{\epsfig{file=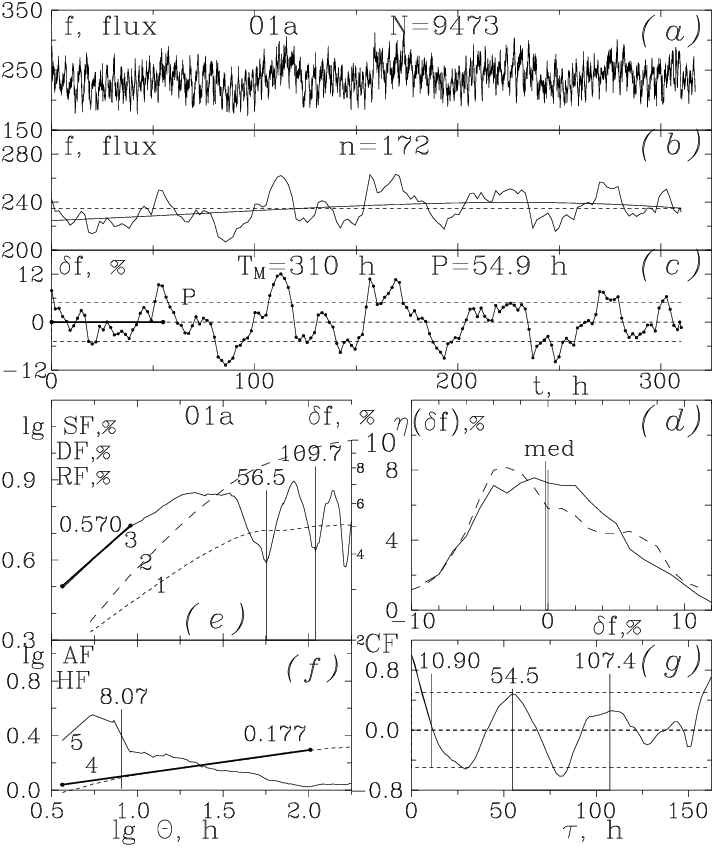, width=.74\textwidth}}  
\caption[] {The processing of LC\#01a. See the text, as well as, Table 1 and 2 in the on-line  Appendix. 
(a): Observed LC with $N$ data points over  time $T$ in hours;
(b): RLC with $n$ data, in fluxes, derived from the observed LC under about 55 times reduction. Dashed and solid lines show the AV  and 3rd degree polynomial fit; 
(c): FLC, derived from the RLC (b) by removal of the polynomial trend and expressed in percents in respect with the polynomial value. Horizontal lines  represent the zero level and $\pm \sigma$ levels. $T_M$ is the observing time and $P$ (solid segment) is the average QP, $P_{12}$;
(d) Histogram of the RLC (dashed line) and FLC (solid line). The vertical lines are the AV and median.; 
(e): DF (1), Eq.~1,  RF (2), Eq.~2. and SF (3), Eq.~3, of the FLC versus the fractal window  $\Theta$ in log-log coordinates. The solid segment shows the SG of the initial part of the SF. The vertical lines  mark two QPs in hours. Note that the right ordinate is graduated in percents; 
(f): HF (1), Eq.~4, and AF (2), Eq.~5, versus the fractal window $\Theta$, in log-log coordinates. The solid segment shows the HG of the middle part of the HF. The vertical line marks th BT of teh AF;
(f): CF (Eq.~6). The vertical lines mark the CT and two QPs in hours. The horizontal lines are the levels of -0.5, 0 and 0.5 of the CF. }
\label{fig.2}   \end{center}  \end{figure*} 

\begin{figure*}
\begin{center} 
\centering{\epsfig{file=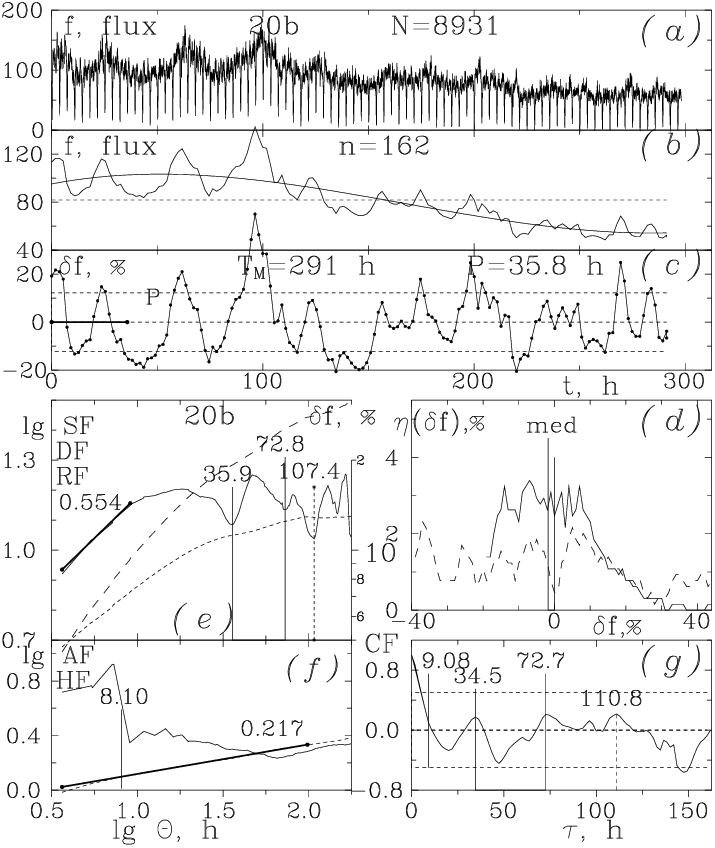, width=.74\textwidth}}
\caption[] {The processing of LC\#20b. See Fig.~2. (a): Observed LC. Vertical segments correspond to eclipses with a period of 
~3.4 h; (b): RLC; (c): FLC; (d): Histograms; (e): DF (1),Eq.~1, RF (2), Eq.~2 and SF (3), Eq.~3, of the FLC versus the fractal window $\Theta$. Here the dashed vertical line marks one additional QP; (f): HF (1), Eq.~4, and AF (2), Eq.~5, versus the fractal window $\Theta$; (f): CF (Eq.~6). The  dashed vertical line marks one additional QP.}
\label{fig.3}   \end{center}  \end{figure*} 

The fractal methods applied here are preferable because they are conceptually simple and they are well applicable at the low signal-to-noise ratio of the QPs. Our   approach was applied for analysis of the flickering of 3 symbiotic stars: RS~Oph, T~CrB, and MWC 560 (hereafter 3SS). The results on them are juxtaposed by Georgiev, Boeva, Stoyanov et al. (2022, hereafter GBS22) and given here in Table 1. 
  
The contents are:

\noindent 1. Light curves and parameters 

\noindent 2. Fractal functions and parameters 

\noindent 3. Quasi-periods and modes 

\noindent 4. Summary

\vspace{0.3cm}
This paper includes also an on-line Appendix, available only in electronic form on the site of the Bulgarian Astronomical Journal, on the page of this paper. It contains 2 tables with data about the LCs and QPs and 14 panels with illustrations. The panels eight LCs \#01a and \#20b are shown here, in Fig.~2 and Fig.~3. 
	
The used abbreviations and subscripts follow: \\ 
3SS -- 3 symbiotic stars in the paper GBS22 \\
AF -- asymmetry (ratio) function, Sect.~2,  Figs.~f; \\ 
AS -- asymmetry (skewness) value, Sect.~1; \\
AV -- average value; \\
BT -- breakdown time point of the AF,  Sect.~2, Figs.~f; \\
CF -- (auto)correlation function, Sect.~2,  Figs.~g;\\
CT -- auto-correlation time. Sect.~2, Figs.~f; \\
DF -- (standard) deviation function, Sect.~2,  Figs.~e; \\
EX -- excess (kurtosis) value, Sect.~1; \\
FLC -- flatten light curve, Sect.~1, Figs.~c; \\
HF -- Hurst function, Sect.~2, Figs.~f; \\
HG -- Hurst gradient, Sect.~2, Figs.~f; \\
LC -- light curve, Sect.~1; \\
QP -- quasi-period in FLC, Sect.~2, Figs.~e, g;\\
RD -- range deviation, Sect.~1, Figs.~e;\\
RF -- range function, Sect.~2, Figs.~e;\\
RLC -- reduced light curve, Sect.~1, Figs.~b; \\
SD -- standard deviation, Sect.~1; \\ 
SF -- structure function, Sect.~2, Figs.~e; \\
SG -- structure gradient, Sect.~2, Figs.~e;\\
av, gr, sd -- average, gradient, standard deviation of the graph. \\

Note that the sub-figures, for example "b", in Fig.~2, Fig.~3, as well as in all panels in the Appendix, will be cited hereafter shortly as "Figs.~b". 

\section*{1. Light curves and parameters} 

Photometric data of AQ~Men with 2 min resolution was obtained from 7 sectors (01, 05, 08, 12, 13, 19 and 20) of the Transiting Exoplanet Survey Satellite (TESS: Ricker et al. 2015). The data are publicly available in the Mikulski Archive for Space Telescopes (MAST)\footnote{http://archive.stsci.edu/}.
More information can be found at the TESS website\footnote{ https://tess.mit.edu/observations/}. In Fig.~1, photometric data in the $g$ band from ASAS-SN (Shappee et al. 2014; Kochanek et al. 2017) from 2018 to 2020 is shown. The shaded areas represent the times of TESS observations. The band-pass of TESS is centered on the classical Ic  filter, but it is wider and spans from 6000  to 10 000, in other words, the telescope observes in the red and the near-infrared wavelengths. 

Each TESS sector includes about 24 days of continuous photometry, with one day pause for data transfer during orbit perigee passage. Each sector contains 2 observed LCs, noted here by "a" and "b". The observations include time intervals after 2015 with BJD 2457000+. The duration of one LC is about 11.5 days or about 300 hours, with about 9000 data points in it (Figs.~a; Appendix, Table 1). 

We aim to analyze the macro-flickering on time scales of tens of hours. Each observed LC is analyzed by using of its derivative forms -- reduced LCs and flattened reduced LCs.

The reduced LCs (RLCs Figs. b) are derived  from the observed LCs (Figs. a) by sliding data average from a window of 110 points. The step of scan is 55 points, i.e. the time resolution is 1.83 ours. Thus the RLC is reduced by about 55 times and contains about 175 data items. RLC under higher or lower  resolution is not smooth enough or too rough, respectively. The RLCs are expressed in fluxes.

The flattened LCs (FLCs, Figs.~c), are derived from the RLCs by removal of a large-scale trend, fitted by a 3rd degree polynomial. The FLCs are deviations with respect to the polynomial level, expressed in percents. The FLCs ensure an uniform approach to parameters and QPs in time series with different sources (Table 1), including e.g. the variations of the Solar Wolf number  (Georgiev, Zamanov, Boeva et al. 2020, hereafter GZB20).

Each RLC or FLC may be characterized by basic statistics: average value (AV), standard deviation (SD), and range deviation (RD). Here these statistics are supplied with the subscripts "0" for the RLC and "1" for the FLC. Note that the AV of the FLC, $AV_1$, is zero by definition. Here and in the previous papers (GZB20; GBS22) the RD is defined to be the half of the peak-to-peak amplitude of the LC (or of any part of the LC). The goal is the use of a RD that is only about 2 times larger than the SD (Figs~e, Fig.~4, Fig.~5).

Other statistics of the scatter of LC deviations are the skewness (AS, asymmetry of the distribution) and the kurtosis (EX, excess, superiority of the tails with  respect to the central part of the distribution). A positive AS corresponds to a heavy positive tail (and vice versa). In comparison with the normal distribution, a positive EX corresponds to a distribution with a light central part and heavy tail(s) (and vice versa). 

In our papers, where we explore long data series and we apply just the definitions for AS and EX, without coefficients for short series. In GZB20. GBS22, and here we also use a modified EX, $EX =\mathrm{lg} (E/3)$, where $E$ is the ordinary kurtosis. Thus the symmetric distributions, such as double exponential, normal, and uniform have $E$ values of 6, 3 and 6/5, as well as $EX$ values 0.3, 0,  and $-0.4$ (Fig.~5c here, and Fig.~5c in GBS22; the uniform distribution is out of the diagram scopes). 

Any LC may be characterized and illustrated by a histogram distribution of its deviations. Here the histograms of the RLCs and FLCs are shown in Figs.~d. The polynomial trends of the RLCs are not strong and the histograms of FLCs are only slightly more narrow.  
 
Figure 4 juxtaposes statistics of the average fluxes and the deviations for all 14 RLCs of AQ~Men. The maximal average flux corresponds to Sector \#05. In Fig.~4 the  low state, noted by I\l{}kiewicz et al. (2021), demonstrated by Sectors \#19 and \#20, shows different statistics than the high state,  containing LCs  \#\# 01, 05, 08, 12, and 13. 

Figure 4a shows that with respect to the high state the RLCs of the low state have lower AVs. Due to the appearance of eclipses with period ~3.4 h, they have higher SDs and Rds. In the low State, the FLCs also show multiple QPs (Section 3). In the majority of the figures the statistics of the low  state recline from the data of the high state.

Figure 4b shows specific particularities of the object: The values of $SD_0$ and $RD_0$ of the high state follow faint linear decreasing  over time connected with the decreasing of the AV.

Figure 4c shows that in high state (excepting \#08b) the SDs and RDs increase with the AV with gradients of 0.018 and 0.021. When \#08b participates in the fits, the gradients are higher, 0.028 and 0.074. Such an increase, but steeper, is a common feature of the 3SS (Table 1). Zamanov, Boeva, Latev, et al. (2016, hereafter ZBL16) investigated 9 interacting binaries, including our 3SS. They derived average gradients of 0.086 and 0.18, respectively. (The second value is noted 2 times lower in respect with the used here definition of RD.) Our gradients $SD_0/AV_0$ and $RD_0/AV_0$ are about 4 and 2 times lower, respectively, than the gradients of ZBL16. However, the number of data is only 10 and the range of the AVs is small for sure conclusions.  

Figure 5 juxtaposes the statistics of the flux deviations only for  the high state using 10 RLCs and FLCs. The deviations in the low state are enlarged by the eclipses, they lie out of the diagrams and they are not used here. 

\begin{figure*}[!htb]  \begin{center} 
\centering{\epsfig{file=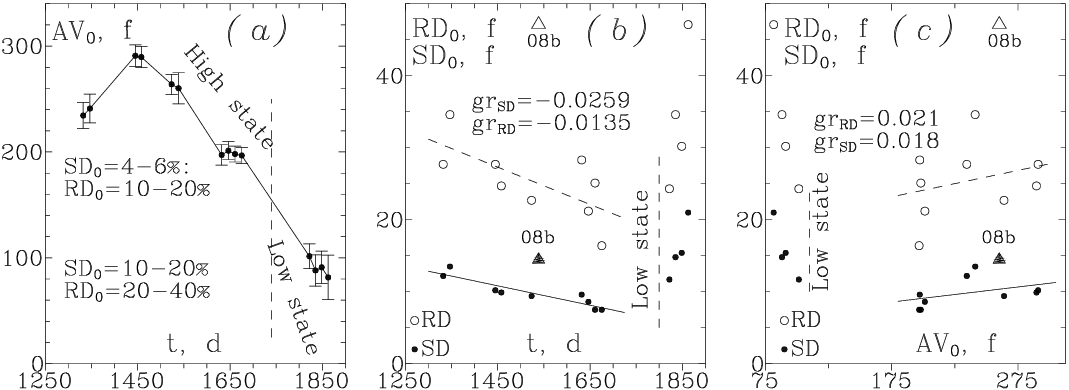, width=.9\textwidth}}
\caption[] {Statistics AV, SD, and RD of the RLCs. 
(a): Behaviour of the average flux $AV_0$ over time $t$, in days, $(d)$ (BJD 2457000+). The error bars show the relevant standard deviations $SD_0$. The scores of the values $SD_0$ and $RD_0$ for high and low states are implemented. 
(b), (c): Behaviour of the deviations $SD_0$ and $RD_0$ from $t$ and $SD_0$. The gradients of the fits for the high state are implemented as $gr_{SD}$ and $gr_{RD}$. The peculiar LC \#08b is not used in the fits.} \label{fig.4}   \end{center}  \end{figure*}

\begin{figure*}[!htb]  \begin{center} 
\centering{\epsfig{file=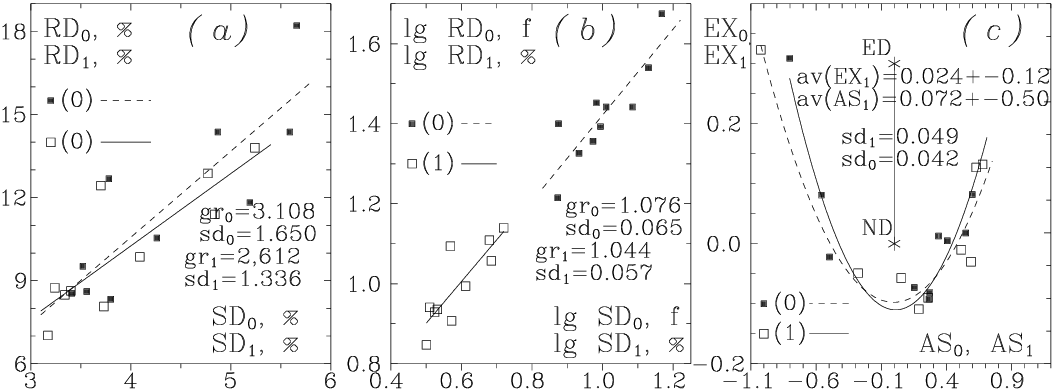, width=.9\textwidth}}
\caption[] {Correlations between statistics of the RLCs and FLCs. (a), (b): Correlations between deviations SDs and RDs of RLCs (0), in fluxes from the AV, and FLCs (1), in percents from the polynomial, in linear and log-log  scales. 
(c): Quadratic dependencies of the modified kurtosis $EX$ on the skewness $AS$ of the RLCs (0) and the FLCs (1). The vertical line contains the positions of the  normal distribution (ND) and the double exponential distribution (2ED). }
 \label{fig.5}  \end{center}  \end{figure*}

Figure 5a shows the correlations between the relative SD and RD in the linear scale. The gradient for FLCs is high, $2.61\pm 1.34$, similar to $2.63\pm 0.90$ for T~CrB (Table 1). 

Figure 5b shows high correlations between the SD and RD from RLCs (0, in fluxes) and FLCs (1, in percents) in log-log scales. The gradients are 1.08 and 1.04 with SDs about 0.06. For 9 flickering binaries ZBL16 derive similar gradient of $1.00\pm 0.08$, but for MWC~560 we found  $0.90\pm 0.05$(GBS22). Our gradients for the  macro-flickering are slightly larger  than 1.0.  

Figure 5c juxtaposes the skewness $AS$ and the modified kurtosis $EX$ of the RLCs and FLCs for the high state only. The data for \#08b  from RLC lie out of the left top corner and they are not used. The EXs obey well pronounced quadratic dependencies like in the case of MWC~560 (GBS22) and in the general case of 21 asymmetric astronomical distributions (Georgiev 2022, Fig.~6a). The average AS and EX are close to zero. Among the already studied 3 SS the data of MWC 560  (GBS22, Table 1) are the most close to these.

\begin{table}   \centering
\caption{Parameters of 3SS and AQ~Men. $R_1/S_1, CC$ -- gradient and correlation coefficient of the relative deviations of the FLCs; AV and SD of the skewness $A$ for the FLCs; AV and SD of the kurtosis $EX$ for the FLCs; AV and SD of the structure gradient $SG$ for the RLCs; AV and SD of the Hurst gradient $HG$ for the FLCs; B and M are the base of the power function and the number of the used modes for the respective regularity model.}  

\begin{tabular}{ccccccccc}      
	\hline\noalign{\smallskip} 
Star & RS~Oph & (B,V)& T~CrB & (U) &  MWC~560 & (B,V) & AQ~Men & flux\\	
	\hline\noalign{\smallskip} 
$RD_1/SD_1, CC$  &   2.43 &  0.90 &  2.63  & 0.90  &   2.15   &  0.93 & 2.61 & 0.80\\
$AS_\mathrm{AV}, AS_\mathrm{SD}$   &   0.08 &  $\pm$0.26 &  0.24 & $\pm$0.16  &   0.13   &  $\pm$0.39 & 0.072 & $\pm$0.50\\
$EX_\mathrm{AV}, EX_\mathrm{SD}$  &   0.17 &  $\pm$0.21 &  0.02  & $\pm$0.19  &  -0.04   &  $\pm$0.11 &  0.024 & $\pm$0.12 \\
$SG_\mathrm{AV}, SG_\mathrm{SD}$ &   0.48 &  $\pm$0.16 &  0.34  & $\pm$0.12  &   0.61   &  $\pm$0.10 & 0.49  &  $\pm$0.14 \\
$HG_\mathrm{AV}, HG_\mathrm{SD}$ &   0.22 &  $\pm$0.06 &  0.18  & $\pm$0.05  &   0.16   &  $\pm$0.07 & 0.198  & $\pm$0.03 \\
$B, M$  & 1.55 &  6    &   2.00    &   5   &
    1.34   &   8   &  1.57     &    7  
 \\                                            
\hline\\  \end{tabular}    \end{table}

\section*{2. Fractal functions and their  parameters} 

Useful fractal functions over the FLC, in log-log scales, reveal specific properties of time series (Figs.~e,~f,~g). 

The FLC, expressed in percents with respect to the polynomial trend, is assumed to be a time series $\delta f(t_n), ~n = 1,2, ... ,N$ (e.g. $N=190$). The fractal characterizing of the FLC is based on a system of scanning with overlapping time windows with sizes $\Theta_j, ~j = 1,2, ... ,J$. The window sizes are designed to be distributed uniformly by lg~$\Theta$. The shortest window is chosen to contain at least 3 data items. The largest window is shown to be applied at least 3 times in the FLCs. The value of $J$ could lie within the range of tens to hundreds.  

Each $j$-th window $\Theta_j$ scans the FLC taking $k = 1,2, ... , K$ different positions. The $k$-th position of the $j$-th window gives fractal indicators,  e.g. $\delta f_\mathrm{SD}(\Theta_j)_k$ and $\delta f_\mathrm{RD}(\Theta_j)_k$. The value of each indicator averaged over all $K$ positions of the $j$-th window, gives relevant fractal parameter, e.g. $<\delta f_\mathrm{SD}(\Theta)>_j$, $<\delta f_\mathrm{RD}(\Theta)>_j$. Hereafter the broken brackets mean averaging of the indicator over all $K$ positions of the window $\Theta_j$. The dependence of the fractal parameter on the window size, in log-log scale, is considered as fractal function. 

The (standard) deviation function (DF) and the range (deviation) function (RF) describe the increase of the respective parameter, characterizing the "roughness" or  "jaggedness" of the FLC with an increase of the window size $\Theta$. The DF and RF are defined as follows:    
\begin{equation}    
DF_j(\Theta)=<\delta f_\mathrm{SD}(\Theta)>_j, \hspace{6mm} 
RF_j(\Theta)=<\delta f_\mathrm{RD}(\Theta)>_j. 
\end{equation}
Figs.~e represent DFs by dashed curves. The RFs are considered to be less interesting.  They are shown here only in Figs.~1 and 2.

\begin{figure}
\begin{center} 
\centering{\epsfig{file=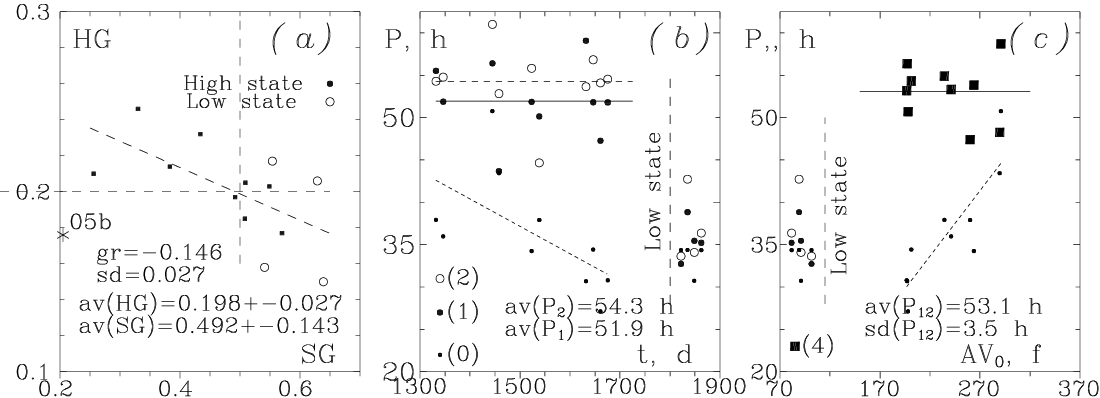, width=0.8 \textwidth}} 
\caption[] {(a): Structure gradients $SG$ versus Hurst gradients $HG$ for all 14 FLCs.  The dashed lines correspond to the "neutral" values $SG=0.5$ and $HG=0.2$.  
(b) and (c): Distributions of the estimated QPs over time $t$ in days and over the average flux $AV_0$. The data from the low state are aside.
The horizontal lines in (b) show the average QP derived from 10 average estimations $P_1$ (1) and 10 average estimations $P_2$ (2).
The horizontal lines in (c) show the average QP with $P=53.1$~h, derived from 20 averages from the high state. } 
\label{fig.6}   \end{center}  
\end{figure} 
\begin{figure} \begin{center} 
\centering{\epsfig{file=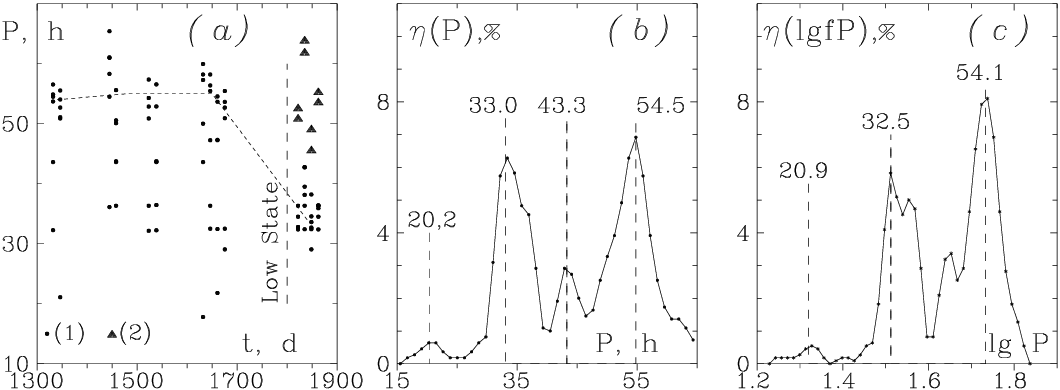, width=.8\textwidth}} 
\caption[] { (a): Distribution of 84 ordinary (1) and 8 additional (2) QPs over  time.  Dashed segments show the trend of the most populated regions of the diagram.    
(b) and (c): Histogram distributions of 92 QPs by linear and logarithmic periods. The 3 QP modes in (c) are assumed to be significant. }
\label{fig.7}   \end{center}  
\end{figure}

When the window size $\Theta$ is small and increases, it incorporates larger FLS structures. The DF and RF increase too. When the largest local structures are exhausted, the DF tends to a plateau, corresponding to the SD of the whole FLC. The RF has higher values and continues to increase, but with decreased rate (Figs.~1e, 2e). An important product of the RF is the rescaled RF or Hurst function (Eq.~3; Figs.~f). 

The structure function (SF) of the RLC is based on a structure indicator -- the absolute difference between the FLC values in the bounds of the window $\Theta$. The value of the FLC at each bound of the window is derived by linear interpolation between the nearest two FLS values (GZB20, Fig.~4a). The average value of this indicator over all its $K$ positions gives the structure parameter  $SF_j$, which is a value of the SF. The structure gradient (SG)  characterizes the slope of the initial quasi-linear part of the SF, here 1/4 of the SF. The SF and SG are defined, as follows:  
\begin{equation}    
SF_j(\Theta)=<\delta f(t+\Theta)-\delta f(t)>_j,   \hspace{6mm} 
SG=\Delta~\mathrm{lg}~(SF)~/~\Delta~\mathrm{lg}~\Theta.    
\end{equation}
Figs.~e represent the SFs by solid curves and the SGs by solid segments. 

While the window $\Theta$ increases, it  incorporates larger structures. The SF increases too, tending to a plateau. However, in contrast with the DF, when  $\Theta$ includes entirely one structure (or 2, 3, ... similar structures together), the quantity of the new larger variations decreases. Then the SF shows one (or more) local minima (Figs.~e). 

The time position $P$ of every minimum of the SF corresponds to a specific size of some structure, i.e. to some quasi-period (QP). In Figs.~e the QPs are marked by vertical lines.  The application of the SF as a QP detector is equivalent to the "phase dispersion minimization technique" of Lafler \& Kinman (1965) (see Ganchev et al. 2017, Fig.~5, GZB20 and GBS22). 

The SG of the SF contains information about the feedback of the flickering (di Clemente et al. 1996; Kawaguchi et al. 1998).  A value $SG<0.5$ corresponds to prior of global flux instabilities and lower variability of the FLC. A value $SG>0.5$ indicates a variability, driven mainly by separate shots, causing high variability of the FLC.  

For example, the cataclysmic variables  KR~Aur (Bachev et al. 2011) and T~CrB (Table 1) have  $SG\approx 0.35$, which gives a hint that the so-called self-organized criticality within an accretion disk may drive the variability of the noise continuum (see Fig.~6a). On the contrary, the flickering of MWC~560, with $SG =0.61$ (Table 1) contains significant contribution of shots of global instability and high variability of the FLC. For AQ~Men we derive the neutral value  $SG_{AV}=0.49 \pm 0.14$. (see Fig.~6a) 

The rescaled RF, known as Hurst function (HF), describes the increase of the ratio $RF_j~/~DF_j$ when the window $\Theta$ increases. The Hurst gradient (HG) is  the slope coefficient of the internal quasi-linear part of the HF, here the internal 1/2 part of the scope. The HF and HG are defined as follows:  
\begin{equation}    
HF_j(\Theta)=<\delta f_{RD}(\Theta)/\delta f_{SD}(\Theta)>_j, \hspace{4mm} 
 HG=\Delta~\mathrm{lg}~(HF)~/~\Delta~\mathrm{lg}~\Theta.    
\end{equation}
Figs.~f represent the HFs in log-log scale  by dashed curves and the HGs by solid segments. Our HFs here and in the cases of the 3SS are smooth and the HGs are well defined. 

Hurst (1951) analyzed the fluctuations in the cumulative discharge of the Nile River and found that it is scaled self-similarly. He defined the rescaled RF by the "peak-to-peak amplitude". We apply the RF by "the half of the peak-to-peak amplitude" (Eq.~1). For this reason, our HG is lower with respect to  the HG of Hurst by $\mathrm lg 2=0.3$. Hurst (1951) derived $HG=0.77$ for the Nile River and our relevant value is $HG=0.47$. 

The HG characterizes the autocorrelation in the LC (Mandelbrot \& Wallis 1969). For our definition a value of $HG<0.2$ indicates a time series with short-term positive autocorrelation, i.e. with higher chaos. Our value of $HG>0.2$ indicates a time series with long-term positive autocorrelation, i.e with lower chaos. For example, the FCs of  MWC~560 (Table 1) has  $HG_{AV}=0.16$, i.e. high chaos. For AQ~Men we derive the neutral value  $HG_{AV}=0.20 \pm 0.03$. (See Fig.~6a.)

Figure 6a represents the distributions of the SGs and HGs of the FLCs. The dashed curve represents a faint anti-correlation. In the cases of RS~Oph and MWC~560 such hints of correlation are subtle (GZB20, GBVS22).  

The asymmetry function (AF) is introduced in GBL20 for any time window $\Theta$ by the empiric dimensionless asymmetry parameter: 
\begin{equation}    
 AF_j(\Theta) = <(\delta f_\mathrm{max} - \delta f_\mathrm{med})~/~(\delta f_\mathrm{med} - \delta f_\mathrm{min})>_j.
\end{equation} 
The AFs are shown in Figs.~e by solid curves. Another AF, defined by the skewness $AS$, has a similar, but more complicated behaviour, especially in the short time scale (GZB20). 

Usually, at large windows, the AF (Eq.~4)   decreases slowly from positive to negative values. However, while the window contains 3--6 adjacent FLC points the AF is clearly positive (Figs.~f). In such time scales the details of the FLC frequently grow with an increasing rate or decline with a decreasing rate (see GZB20, Fig.~4c and Fig.~10). In the majority of the FLCs the AF poses a breakdown time-point (BT) at some short time window. The BTs are marked by vertical lines in Figs.~e. The BT of the AF appears to be an estimator of the half-period of the shortest peaks or quarter-period QP-like fluctuations of the LC: $P_{BT}=4\times BT$.

The autocorrelation function (CF) is useful for FLCs with constant cadence. The CF characterizes the change of the mutual correlation of the FLC values on dependence of the time lag (shift) $\tau $: 
\begin{equation}    
CF_j(\tau) = <\delta f(t+\tau) \times \delta f(t)>_j. 
\end{equation} 
The CF is shown in Figs.~g. By definition $CF(0)=1$. When $\tau $ is small and increases, the CF decreases. It is characterized by the CF time lag correlation time (CT), $CT=\tau_0$, at which the CF takes firstly the zero value. The CT of the CF appears to be an estimator of the quarter-period of the shortest QP-like fluctuations of the FLC: $P_{CT}=4\times CT$. At larger $\tau$, the CF fluctuates. The positions of the maxima of the CF correspond to QPs $P, 2P, 3P...$. 

Figs.~g show CFs with their CTs and QPs,  marked by vertical lines. The CF maxima, that confirm the QPs, detected by the SF minima, are marked by vertical lines like in the SF in Figs.~e.

\section*{3. Quasi-periods and their modes} 

The periods $P$ of the QPs are detected by the minima of SFs (Eq.~2) and the respective maxima of the CFs (Eq.~5). The two shortest  of the dominating QPs,  with times $P_1$ and $P_2\approx 2P_1$, are marked in Figs.~e and g by solid vertical lines. We use half of $2P$ as an estimator of the main QP. We use also the BT of the AF and CT of the CF as indirect estimators of  1/4 of the main QPs. So, every FLC gives us 6 estimations of the QPs and we use them by 3 pairs. 

In Figs.~6b and 6c, we show 3 kinds of QP  estimations: $P_1$, $P_2$, and $P_0$. We account the averages of the shortest well pronounced QPs from SFs and CFs as the basic ones, with $P_1=(P_{1SF}+P_{1CF})/2$. We consider the averages of the next coming QPs (about the expected positions $2P$) to be $P_2=(P_{2SF}+P_{2CF})/4$.

\begin{table}   
\centering
\caption{List of the QP modes. $P_M$ -- model prognoses of QP (in hours); 
$P_O$ --  observed QP; source of data: G -- this work, A -- Armstrong et al. (2013), 
B -- Burch (2021), and C -- Chen et al. (2001).} 
\begin{tabular}{ccccccc}      
	\hline\noalign{\smallskip} 
& Mode & $P_M$,~h & lg~$P_M$ & $P_O$,~h & lg~$P_O$    & Source  \\
	\hline\noalign{\smallskip} 
	 & -1 &   2.28 &   0.3576 &    -    &    -    &  -      \\
	 &  0 &   3.58 &   0.553 &   3.47  &  0.540  &  A      \\
	 &  1 &   5.61 &   0.749 &    -    &    -    &   -     \\
	 &  2 &   8.81 &   0.945 &   9.125 &  0.960  &  A      \\
	 &  3 &  13.82 &   1.142 &     -   &    -    &   -     \\
	 &  4 &  21.69 &   1.336 &   20.9  &  1.32   &  G      \\
	 &  5 &  34.04 &   1.532 &   32.5  &  1.512  &  G      \\
	 &  6 &  53.42 &   1.728 &   54.1  &  1.734  &  G      \\
	 &  7 &  83.83 &   1.923 &   91.255&  1.960  &  C      \\
	 &  8 & 131.56 &   2.119 &    -    &     -   &   -    \\
	 &  9 & 206.46 &   2.315 &   198.0 &  2.297  &  B     \\
	 & 10 & 324.02 &   2.511 &    -    &    -    & -      
\\
\hline\\  
\end{tabular}    
\end{table}

We account also shorter QPs, estimated by the BTs and CTs as $P_0=(4P_{BT}+4p_{CT})/2$. These QPs are shown in Figs.~6b and 6c. Our 10 FLCs from the high state give 12 QPs of the type $P_1$ and $P_2$. They are shown around the horizontal lines. Other 10 QPs of the type $P_0$ are shown around the sloped lines. The FLCs of the low stage give other 12 similar QPs, shown aside. 

Figs.~6b and 6c show 30+12 averaged estimations of QPs, distributed over time and AVs. The estimations $P_1$ and $P_2$ from the high state have separate averages of 51.9~h and 54.3~h, and together their  average is  $P_{12}=53.1\pm 6.6\%$~h. In the high state, the value of the main QP seems to be approximately constant. The values of $P_0$ show some linear trends giving evidence that towards the low state the values of the shortest QPs increase.

For general revealing of the QP modes, we inventorize all $14\times 2\times 3=84$~QPs. Four FLCs of the low State (sectors \#19 and \#20) have a more complicated structure. The 4 SFs and 4 CFs of the low state show one additional QP in every case, i.e, additional $4\times 2=8$ QPs. These QPs are marked in Figs.~e and g by dashed vertical lines. They are added to the common list.

Figure 7a shows all 92 QPs over time. Note that the most populated regions switch from about $P=55$~h in the high state to about $P=33$~h in the low state. The additionally detected 8 QPs (triangles) appear at $P>45$~h.

Figures 7b and 7c show the (histogram) distributions of 92 QPs over linear and log scale of the periods. The QPs show preferable values, with modes $M$, of 21, 33, 43, and 54 h. The modes at 20 h and 43 h are poorly pronounced. However, the general picture of the observed modes (Fig.~8) forces us to ignore the QPs at about 43 h. The QPs at about 54.5 h corresponds to the superorbital period at 57 h of I\l{}kiewicz et al. (2021).

Figure 8 illustrates the general regularity (near-commensurabiliy) of the QP modes of AQ~Men. The fit is based on our 3 QP modes mentioned above (Fig.~7c) plus 4 modes from literature (Table 2). The orbital period of AQ~Men is estimated via the eclipses in Sectors \#19 and \#20 to be ~3.4 h. Estimations are published by Chen et al. (2001) -- 3.12 h, Armstrong et al. (2013) -- 3.39 h and 3.47 h, and Bruch (2021) -- 3.91 h. The average value  is $3.4 \pm 0.2$~h. This period is assumed to correspond to mode M=0. Then the regularity of the modes is consistent with the following models:
\begin{equation}    
\mathrm{lg}~P = 0.553 + 0.196 \times M, 
  \hspace{6mm} \mathrm{or} \hspace{6mm} 
P_M = 3.595\times 1.568 ^M. 
\end{equation} 
The prognosed orbital period (for M=0) is 3.595 h with SD 0.027 h or 6.4\%. The regularity parameters (bases of the power functions) of the 3SS and AQ~Men are displayed in Table 1. For unknown reasons (i) these regularities appear and (ii) the bases of their power functions are various.

\begin{figure}  \begin{center} 
\centering{\epsfig{file=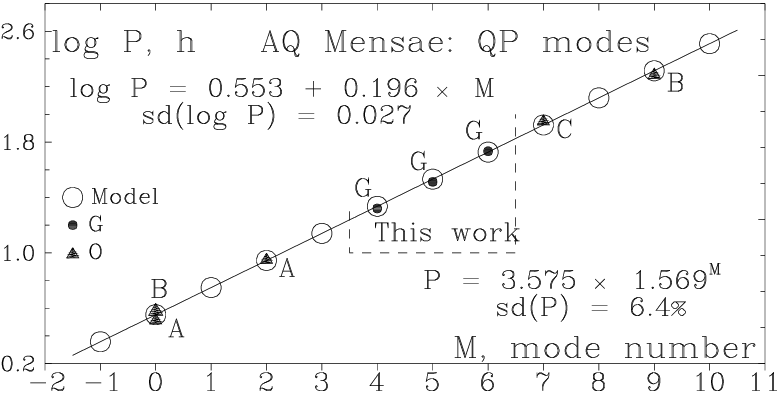, width=.7\textwidth}} 
\caption[] {The regularity of 7 available QP modes of AQ~Men: ($\bigcirc$) -- prognoses from the derived power model,(G) -- from this work. The ones from the literature include: (A) -- from Armstrong et al. (2013), (B) -- from Bruch (2020), and (c) -- from Chen et al. (2001). The  linear and power models are implemented. }
\label{fig.8}   \end{center}  \end{figure}

Table 2 contains the list of the QP modes, shown in Fig.~8. Three internal and two external positions are empty. Our model predicts QP modes at 2.28, 5.61, 13.82, 131.56, and 324.02 h.

\section*{Summary}

$\hspace {6mm}$1. In the high state of AQ~Men the gradients $RD_0/AV_0$ and $SD_0/AV_0$ (Fig.~4c) are about 4 and 2 times lower than for the 9 interacting binaries of ZBL16. From this point of view, the macro-flickering of AQ~Men is different from the stars in ZBL16. Maybe the number of data here (10 LCs) is low for  sure conclusions.  

2. In the high state of AQ~Men the gradient $RD_1/SD_1$ of the FLCs is high (Fig.~5a), similar to the gradients of T~CrB (Table 1). This is evidence for a relatively high contribution of large deviations in the LCs. Simultaneously, the dimensionless gradients lg~$RD_0$/lg~$SD_0$ (from the RLCs) and lg~$RD_1$/lg~$SD_1$ (from the FLCs) in Figure~5b are a little larger than unity, like in the 9 variables of ZBL16. From this point of view, the macro-flickering in the high state of AQ~Men is similar to the flickering of the interacting binaries. The gradient $RD_1/SD_1$ in Figure~5b seems to be a universal dimensionless characteristic of the flickering. This gradient may be regarded as a relation of the type "jaggedness-to-roughness" of the time series.   

3. The average values of the skewness (AS) and the modified kurtosis (EX) of the macro-flickering of AQ~Men are close to zero (Fig.~5c). Among the 3SS the most similar data belongs to MWC~560 (Table 1). The kurtosis (EX) obey well-defined quadratic dependence on the skewness (Fig.~5c). Such dependences are found for MWC~560 (GBS22), as well as in the general case of 21 asymmetric astronomical distributions (Georgiev, 2022). 

4. The average SG and HG of the FLCs in Figure~6a are 0.5 and 0.2, respectively. These neutral values do not bring additional specific information. Otherwise, Figure~6a shows a hint of anti-correlation between SG and HG, like in the cases of RS~Oph and MWC~560 (GZB20, GBVS22).   

5. We reveal 92 QPs in the range 20--70 h. The most populated 3 QP modes correspond to 20.9, 32.5 and 54.1 h or 1.149, 0.738, and 0.444 c/d (Fig.~7c). We consider that the last of them corresponds to the superorbital period, derived of I\l{}kiewicz et al. (2021) to be 57.02 h. Other 4 QP modes are found in the literature. The regularity of these 7 QP modes obeys a power function with a base 1.57 with SD of 6.4.\% (Fig.~8). Three internal and two external possible modes are prognosticated. The regularity bases and mode numbers for the 3SS and AQ~Men are shown in Table 1. 

{\bf Acknowledgments: }
This work is supported by the grant  K$\Pi$-06-H28/2 08.12.2018 of the Bulgarian National Science Fund. This paper includes data collected with the $TESS$ mission, obtained from the MAST data archive at the Space Telescope Science Institute (STScI). Funding for the $TESS$ mission is provided by the NASA's Science Mission Directorate. STScI is operated by the Association of Universities for Research in Astronomy, Inc, under NASA contract NAS 5-26555.  

{}

\clearpage 

\begin{appendix}

{\bf APPENDIX:  The Appendix contains  2 tables and 14 panels. } 


\begin{table}   \centering
\caption{\textbf{Basic data about 14 light curves:} 
1~-- designation of the LC,  
2~-- middle moment of the LC (d),
3~-- duration of the LC (min), 
4~-- number of data in the FLC,  
5~-- cadence (min),
6~-- AV (flux), 
7~-- SD$_O$ of the RLC (\%),
8~-- RD$_O$ of the RLC (\%),
9~-- SD$_1$ of the FLC (\%),
10~-- RD$_1$ of the FLC (\%). }
	\begin{tabular}{cccccccccc}     
\hline\noalign{\smallskip} 
\#LC & $T_a$ & $T_M$ & $n$ & $dt$ & $AV$   &  sd0\% &  rd0\% &  sd1\% &  rd1\% \\ 
\hline\noalign{\smallskip} 
 1 & 2 & 3 & 4 & 5 & 6 & 7 & 8 &  9  & 10  \\
\hline\noalign{\smallskip}
 01a & 1331.90 & 310.0 & 172 & 1.813 & 234.4 &  5.19 & 11.82 &  4.92 & 11.39 \\
 01b & 1346.44 & 316.7 & 175 & 1.820 & 241.1 &  5.59 & 14.37 &  5.30 & 13.80 \\
 05a & 1444.89 & 286.1 & 159 & 1.811 & 290.9 &  3.52 &  9.52 &  3.40 &  8.49 \\
 05b & 1457.91 & 297.6 & 164 & 1.826 & 289.6 &  3.41 &  8.54 &  3.20 &  7.02 \\
 08a & 1523.21 & 274.6 & 152 & 1.818 & 263.9 &  3.56 &  8.61 &  3.28 &  8.74 \\
 08b & 1538.50 & 162.7 &  91 & 1.808 & 260.1 &  5.66 & 18.21 &  3.00 &  7.34 \\
 12a & 1631.97 & 330.2 & 183 & 1.814 & 197.2 &  4.87 & 14.37 &  4.97 & 13.59 \\
 12B & 1646.46 & 301.9 & 168 & 1.808 & 201.3 &  4.26 & 10.54 &  4.07 & 10.31 \\
 13a & 1660.80 & 323.8 & 180 & 1.809 & 198.0 &  3.78 & 12.68 &  3.73 & 12.62 \\
 13B & 1675.49 & 322.8 & 179 & 1.814 & 196.7 &  3.80 &  8.33 &  3.78 &  8.26 \\
 19a & 1822.04 & 279.0 & 155 & 1.812 & 101.5 & 11.51 & 23.93 &  9.40 & 20.44 \\
 19B & 1835.08 & 285.2 & 158 & 1.817 &  89.1 & 16.77 & 39.26 & 15.21 & 33.23 \\
 20a & 1848.65 & 288.5 & 160 & 1.815 &  91.0 & 16.94 & 33.22 &  9.66 & 23.16 \\
 20b & 1862.61 & 291.2 & 162 & 1.809 &  81.6 & 25.75 & 57.78 & 12.58 & 32.82 
\\
 \hline\\   \end{tabular}     \end{table}

\newpage
\begin{table}   \centering
\caption{\textbf{Additional data about 14 light curves:}
1~-- designation of the LC,  
2~-- AS$_O$ of the RLC (\%),
3~-- EX$_O$ of the RLC (\%),
4~-- AS$_1$ of the FLC (\%),
5~-- EX$_1$ of the FLC (\%),
6~-- SG of the FLC (\%),
7~-- HG of the FLC (\%),
8--11 -- QPs (h); See Section 3 of the paper text.}
	\begin{tabular}{ccccccccccc}     
\hline\noalign{\smallskip} 
\# & $AS_0$ & $EX_0$ & $AS_1$ & $EX_1$ &  $SG$ & $HG$ & $Q_0$ & $Q_1$ & $Q_2$ & $Q_{12}$ \\
\hline\noalign{\smallskip}  b
1 & 2 & 3 & 4 & 5 & 6 & 7 & 8 & 9 & 10 & 11 \\
\hline\noalign{\smallskip}
 01a &  0.189 & -0.109 &  0.214 & -0.061 & 0.570 & 0.177 & 29.1 & 55.5 & 54.3 &  54.9 \\
 01b & -0.278 & -0.049 & -0.469 & -0.021 & 0.508 & 0.185 & 30.7 & 51.1 & 54.8 &  53.3 \\
 05a &  0.627 &  0.127 &  0.384 &  0.024 & 0.383 & 0.214 & 41.7 & 56.4 & 61.0 &  58.7 \\
 05b &  0.053 & -0.057 &  0.268 & -0.091 & 0.305 & 0.176 & 30.8 & 43.7 & 52.8 &  48.2 \\
 08a &  0.588 & -0.030 &  0.404 &  0.001 & 0.509 & 0.205 & 26.2 & 51.9 & 55.8 &  53.8 \\
 08b & -1.880 &  0.468 & -0.386 &  0.047 & 0.256 & 0.210 & 29.9 & 50.1 & 44.7 &  47.4 \\
 12a &  0.676 &  0.132 &  0.663 &  0.091 & 0.482 & 0.197 & 26.2 & 59.1 & 53.6 &  56.4 \\
 12B &  0.513 & -0.010 &  0.569 &  0.027 & 0.434 & 0.232 & 26.3 & 51.8 & 56.8 &  54.3 \\
 13a & -1.023 &  0.323 & -0.785 &  0.306 & 0.330 & 0.246 & 19.0 & 47.3 & 54.1 &  50.7 \\
 13B &  0.259 & -0.090 &  0.278 & -0.081 & 0.539 & 0.203 & 22.7 & 51.8 & 54.5 &  53.2 \\
 19a & -0.232 & -0.144 & -0.164 & -0.116 & 0.629 & 0.206 & 26.3 & 32.7 & 33.6 &  33.2 \\
 19B &  0.453 & -0.027 &  0.425 & -0.082 & 0.639 & 0.150 & 26.3 & 38.8 & 42.7 &  40.8 \\
 20a & -0.420 & -0.146 & -0.141 & -0.071 & 0.541 & 0.158 & 22.6 & 35.5 & 34.1 &  34.8 \\
 20b &  0.406 & -0.075 &  0.703 &  0.055 & 0.554 & 0.217 & 26.3 & 35.2 & 36.4 &  35.8 
 \\
\hline\\   \end{tabular}     \end{table}
   
\newpage

\begin{figure*}[!htb]  \begin{center} 
\centering{\epsfig{file=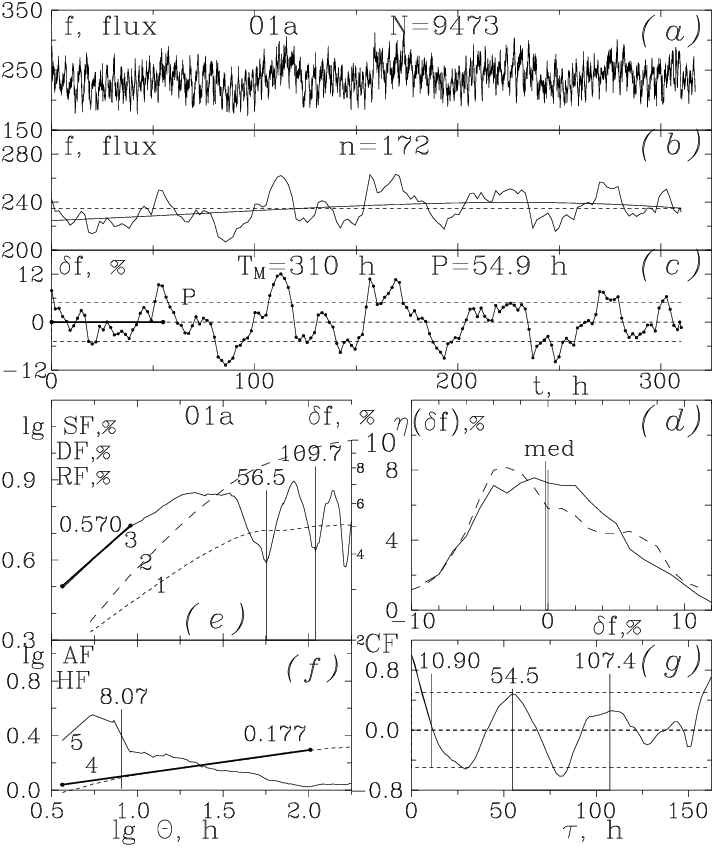, width=.44\textwidth}}
\centering{\epsfig{file=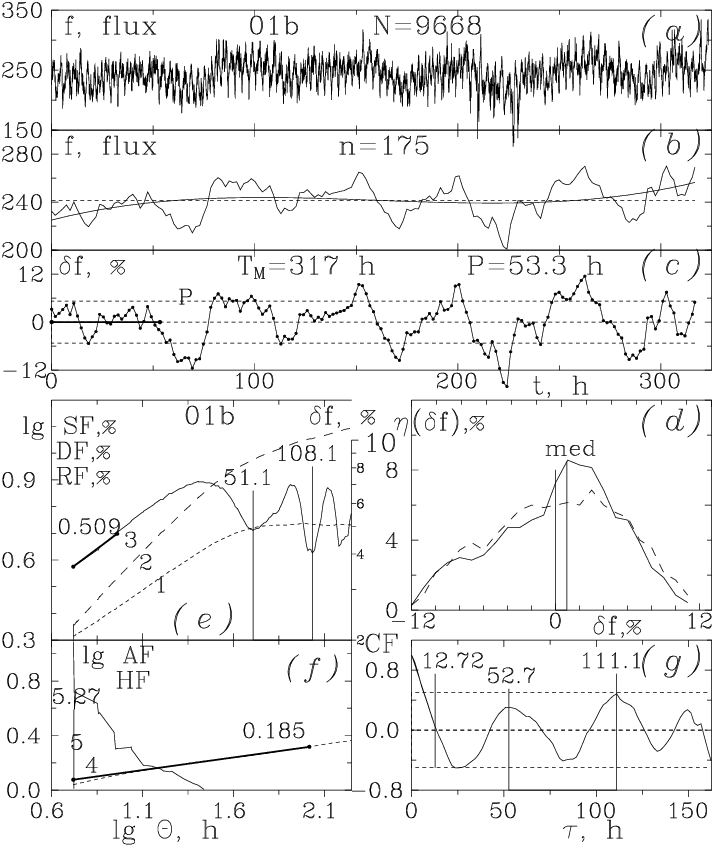, width=.44\textwidth}}
  \caption[] {Sector 1, light curves 01a, 01b.  See Figs.~2 and 3 in the paper.}
\label{fig.01}   \end{center}  \end{figure*} 

\begin{figure*}[!htb]  \begin{center} 
\centering{\epsfig{file=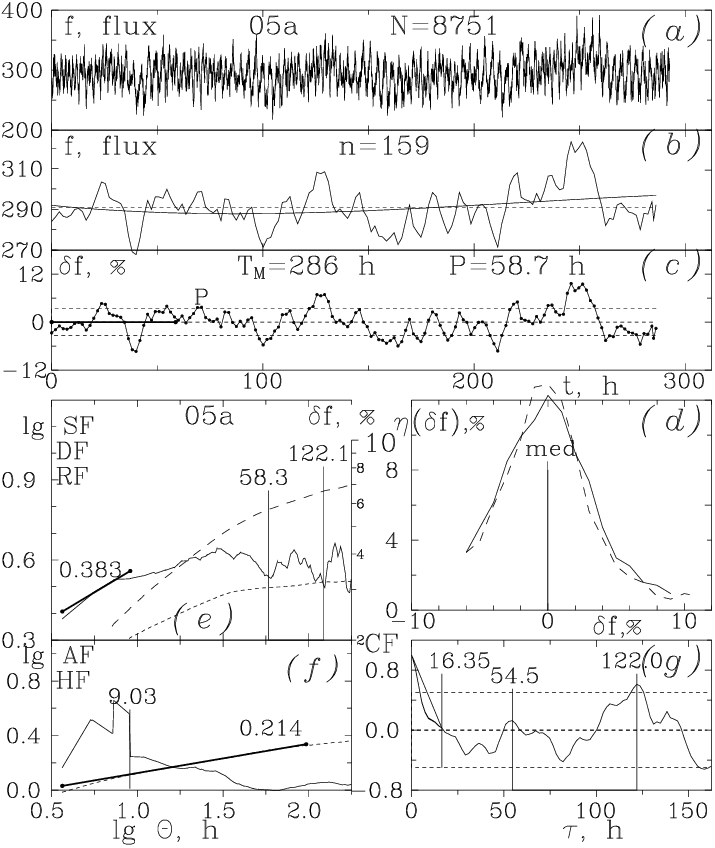, width=.44\textwidth}}
\centering{\epsfig{file=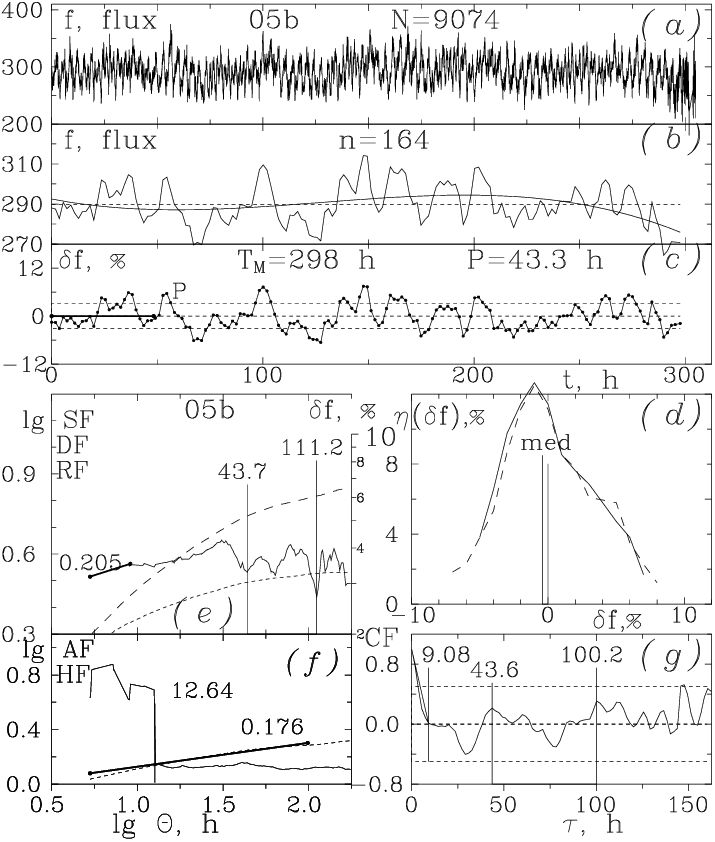, width=.44\textwidth}}
\caption[] {Sector 5, light curves 05a, 05b.}
\label{fig.02}   \end{center}  \end{figure*} 

\begin{figure*}[!htb]  \begin{center} 
\centering{\epsfig{file=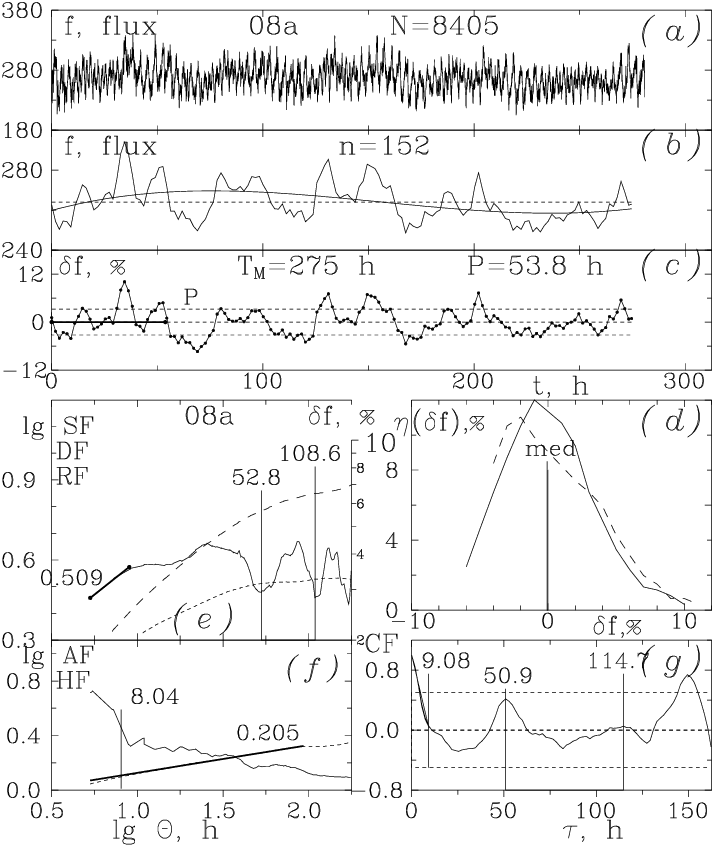, width=.44\textwidth}}
\centering{\epsfig{file=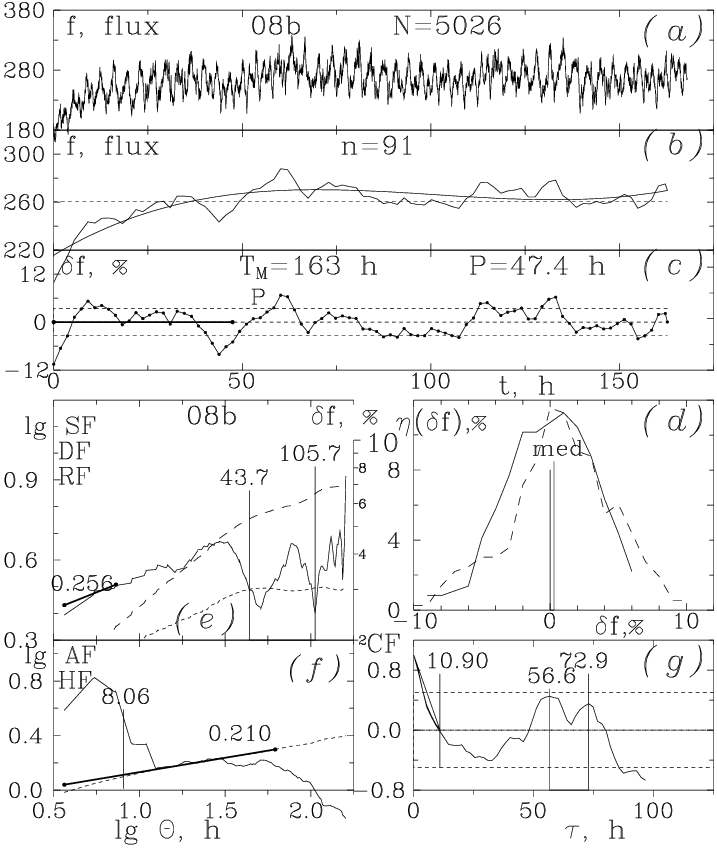, width=.44\textwidth}}
\caption[] {Sector 8, light curves 08a, 08b.}
\label{fig.03}   \end{center}  \end{figure*} 

\begin{figure*}[!htb]  \begin{center} 
\centering{\epsfig{file=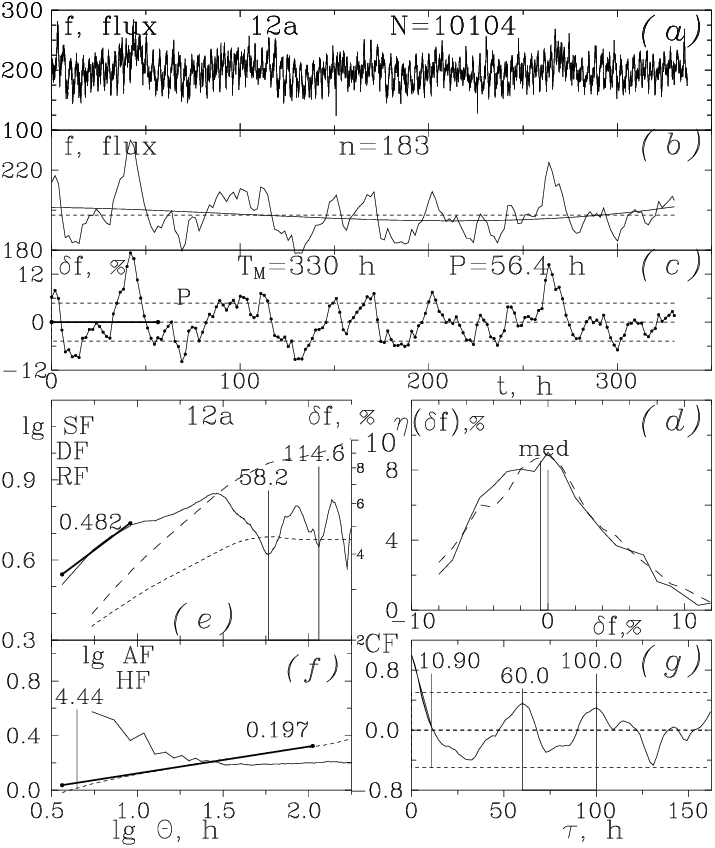, width=.44\textwidth}}
\centering{\epsfig{file=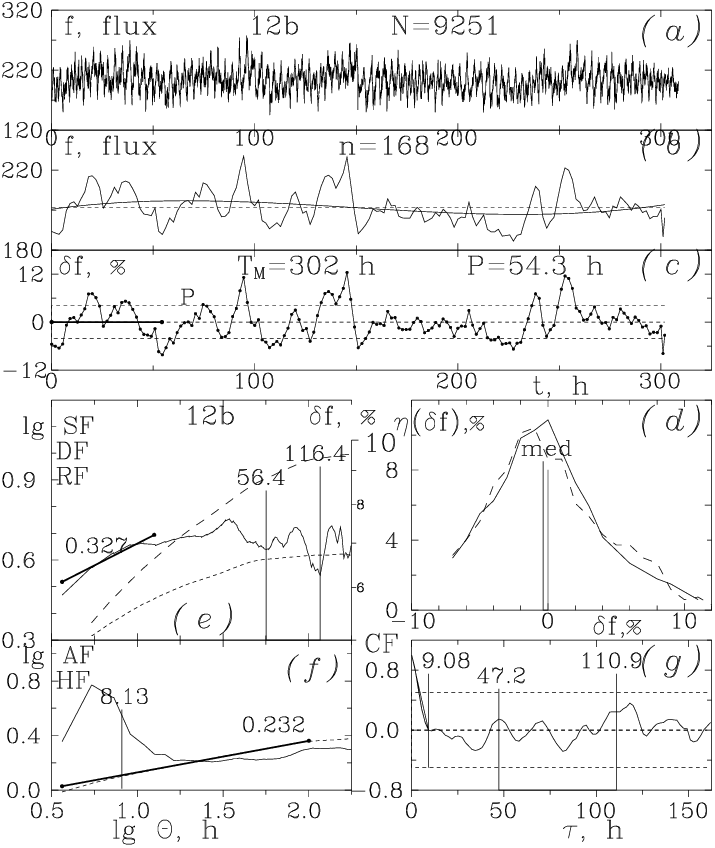, width=.44\textwidth}}
\caption[] {Sector 12, light curves 12a, 12b.}
\label{fig.04}   \end{center}  \end{figure*} 

\begin{figure*}[!htb]  \begin{center} 
\centering{\epsfig{file=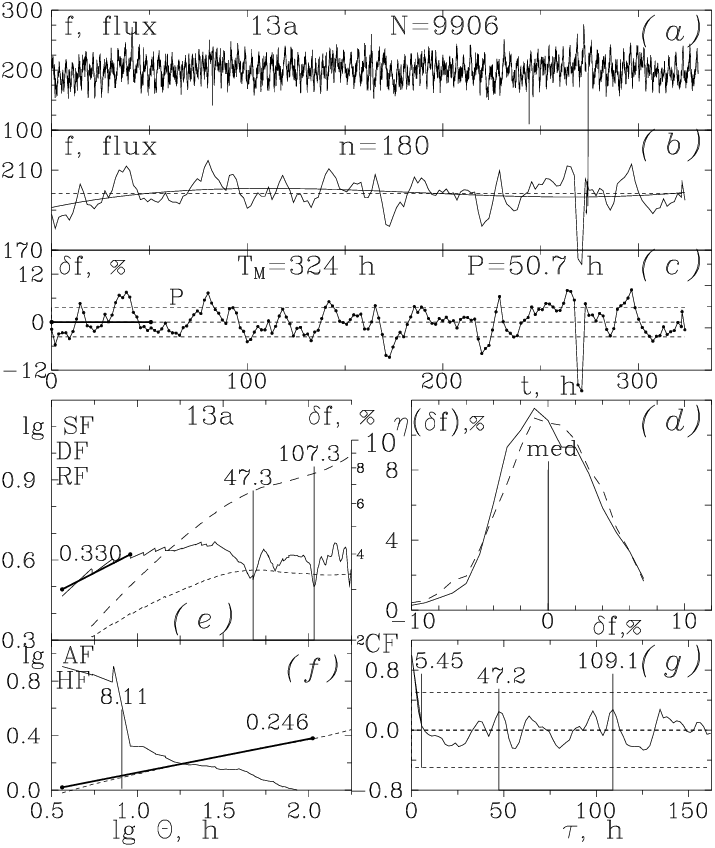, width=.44\textwidth}}
\centering{\epsfig{file=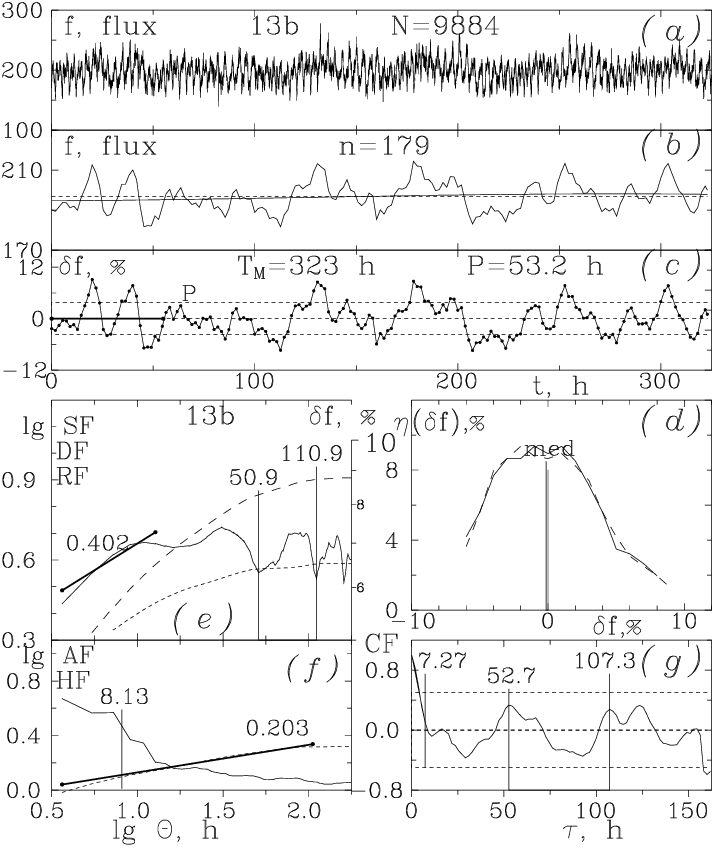, width=.44\textwidth}}
\caption[] {Sector 13, light curves 13a, 13b.}
\label{fig.05}   \end{center}  \end{figure*} 

\begin{figure*}[!htb]  \begin{center} 
\centering{\epsfig{file=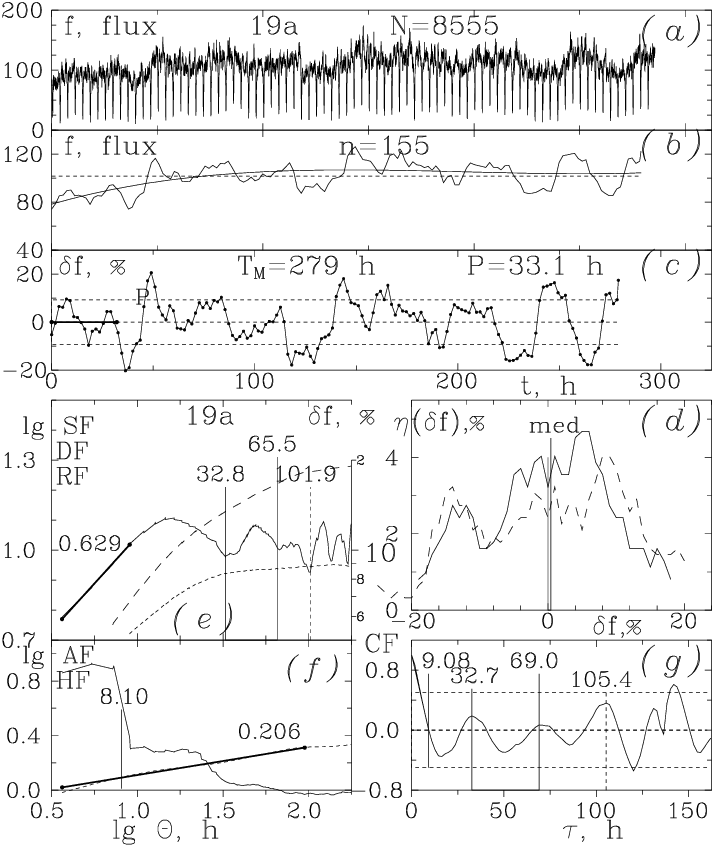, width=.44\textwidth}}
\centering{\epsfig{file=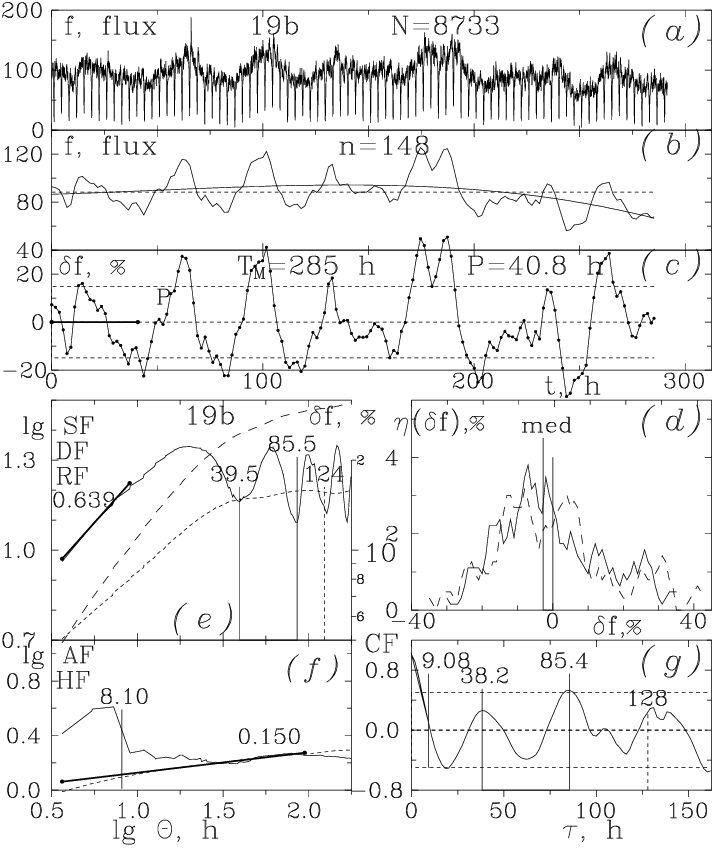, width=.44\textwidth}}
\caption[] {Sector 19, light curves 19a, 19b.}
\label{fig.06}   \end{center}  \end{figure*} 

\begin{figure*}[!htb]  \begin{center} 
\centering{\epsfig{file=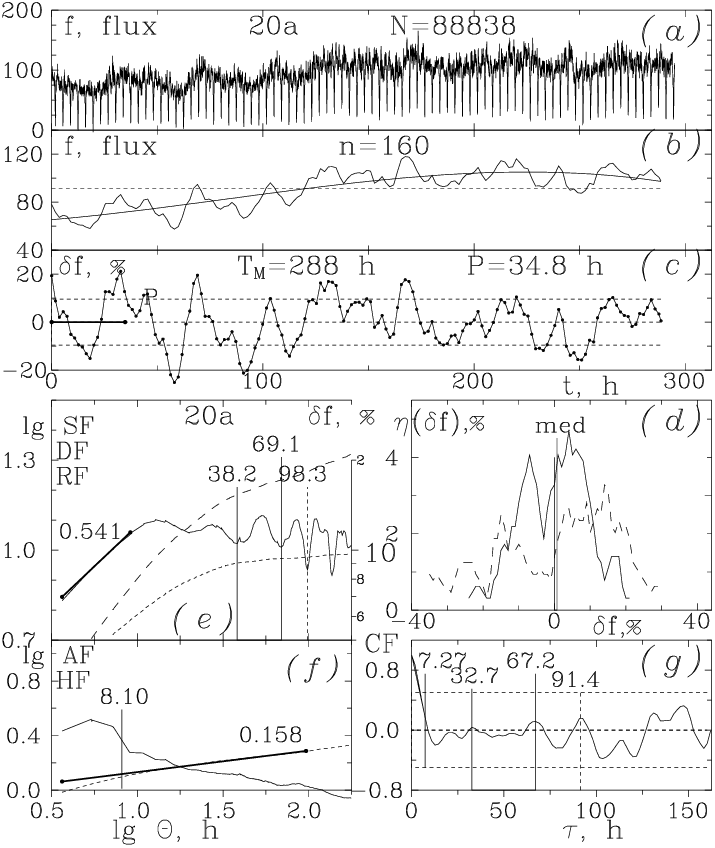, width=.44\textwidth}}
\centering{\epsfig{file=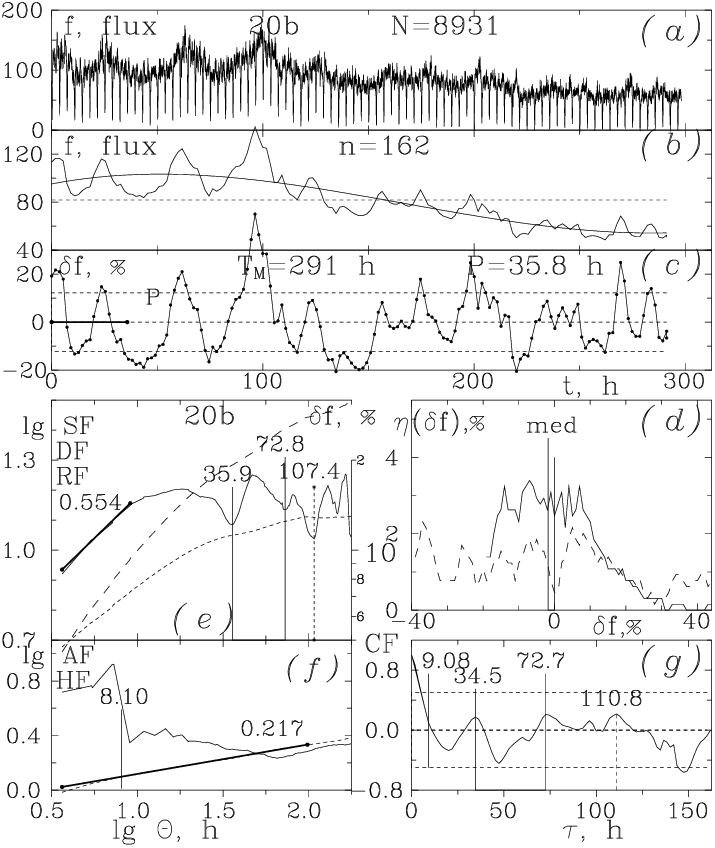, width=.44\textwidth}}
\caption[] {Sector 20, light curves 20a, 20b.}
\label{fig.07}   \end{center}  \end{figure*} 

\end{appendix}

\end{document}